\documentclass[12pt]{article}

\usepackage[T1]{fontenc} 				
\usepackage{microtype} 					

\usepackage{amssymb,amsmath,amscd}
\usepackage[english]{babel}
\usepackage{titlesec}
\usepackage{slashed}
\usepackage{hyperref}
\usepackage{graphicx,xcolor}
\usepackage[margin=0.5cm]{caption}

\usepackage{geometry} 					
\numberwithin{equation}{section} 		

\titleformat{\section}[block]{\Large\bfseries\centering}{\thesection}{1em}{} 
\titleformat{\subsection}[block]{\bfseries}{\thesubsection}{1em}{} 
\titlespacing*{\section}{0pt}{1em}{1em}

\hypersetup{
	colorlinks=true,
	linkcolor=dark-blue,
	citecolor=dark-red,
	urlcolor=dark-green,
	linktoc=all
}

\graphicspath{{./figures/}}

\newcommand\Tstrut{\rule{0pt}{3ex}}         

\definecolor{dark-gray}{gray}{0.20}
\definecolor{gray}{gray}{0.30}
\definecolor{light-gray}{gray}{0.80}
\definecolor{dark-red}{rgb}{0.7,0,0}
\definecolor{dark-green}{rgb}{0.1,0.4,0}
\definecolor{dark-blue}{rgb}{0.3,0.3,0.7}
\definecolor{light-blue}{rgb}{0.8,0.8,1}
\definecolor{cardinal}{rgb}{0.6,0,0}
\definecolor{darkgreen}{rgb}{0,0.5,0}
\definecolor{golden}{rgb}{0.92, 0.7, 0}
\definecolor{midnight}{rgb}{0, 0, 0.5}
\definecolor{darkblue}{rgb}{0.2, 0, 0.8}
\definecolor{forestgreen}{rgb}{0.13, 0.55, 0.13}

\newcommand\cK{\mathcal{K}}

\newcommand\cM{\mathcal{M}}
\newcommand\cN{\mathcal{N}}

\newcommand\cR{\mathcal{R}}

\newcommand{\dd}{\mathrm{d}}
\newcommand{\e}{\mathrm{e}}

\renewcommand\Re{{\rm Re}} 
\renewcommand\Im{{\rm Im}}
\newcommand{\dvol}{{\rm vol}}
\newcommand{\vol}{\mathbf{V}}

\newcommand{\f}[2]{\frac{#1}{#2}}

\newcommand{\nn}{\nonumber}
\newcommand{\ds}{{\rm d}s}
\newcommand{\wti}[1]{\widetilde{#1}}
\newcommand{\ti}[1]{\tilde{#1}}
\newcommand{\AdS}{{\rm AdS}}
\newcommand{\dS}{{\rm dS}}

\newcommand\SO{\mathrm{SO}}
\newcommand\ISO{\mathrm{ISO}}

\newcommand\SU{\mathrm{SU}}

\title{\fontsize{20pt}{24pt}\selectfont\textbf{Bubble instability of mIIA on $\mathrm{AdS}_4\times S^6$}\vspace{3mm}}

\author{Pieter Bomans$^{\bullet\circ\dagger}$, Davide Cassani$^{\circ}$,\\
	Giuseppe Dibitetto$^{\bullet\circ}$ and Nicol\`o Petri$^\ddagger$\\[5mm]
	\normalsize $^{\bullet}$Universit\`a di Padova, Dipartimento di Fisica e Astronomia\\
	\normalsize via Marzolo 8, 35131 Padova, Italy\\[2mm]
	\normalsize $^{\circ}$INFN, Sezione di Padova\\
	\normalsize via Marzolo 8, 35131 Padova, Italy\\[2mm]
	\normalsize $^{\dagger}$Joseph Henry Laboratories, Princeton University\\
	\normalsize Princeton, NJ 08544, U.S.A.\\[2mm]
	\normalsize $^{\ddagger}$University of Oviedo, Department of Physics\\
	\normalsize Avda. Federico Garcia Lorca s/n, 33007 Oviedo, Spain\\[2mm]
	\texttt{\small\textcolor{dark-green}{$\big\{$\href{mailto:pieter.bomans@pd.infn.it}{pieter.bomans},\href{mailto:davide.cassani@pd.infn.it}{davide.cassani},\href{mailto:giuseppe.dibitetto@pd.infn.it}{giuseppe.dibitetto}$\big\}$@pd.infn.it}}\\
	\texttt{\small\textcolor{dark-green}{\href{mailto:petrinicolo@uniovi.es}{petrinicolo@uniovi.es}}}
}

\date{}

\begin{document}  
	
\maketitle
	
\begin{abstract}
\noindent We consider compactifications of massive IIA supergravity on a six-sphere. This setup is known to give rise to non-supersymmetric AdS$_4$ vacua preserving $\SO(7)$ as well as G$_2$ residual symmetry. Both solutions have a round $S^6$ metric and are supported by the Romans' mass and internal $F_{6}$ flux. While the $\SO(7)$ invariant vacuum is known to be perturbatively unstable, the G$_2$ invariant one has been found to have a fully stable Kaluza-Klein spectrum. Moreover, it has been shown to be protected against brane-jet instabilities. Motivated by these results, we study possible bubbling solutions connected to the G$_2$ vacuum, representing non-perturbative instabilities of the latter. We indeed find an instability channel represented by the nucleation of a bubble of nothing dressed up with a homogeneous D2 brane charge distribution in the internal space. Our solution generalizes to the case where $S^6$ is replaced by any six-dimensional nearly-K\"ahler manifold.
\end{abstract}
	
\clearpage
	
{
	\hypersetup{linkcolor=black}
	\tableofcontents
}
	
\newpage	
	
\section{Introduction}

The microscopic interpretation in string theory of classical gravitational solutions without supersymmetry is one of the main challenges in high-energy physics. From a top-down perspective, the need for a thorough understanding of the possible supersymmetry breaking mechanisms has been manifest since the very early days of superstring theory, motivated by the demand of constructing phenomenologically realistic models. More recently, the Swampland program~\cite{Vafa:2005ui} has emphasized the importance of taking a bottom-up approach and asking what subset of all possible candidate low-energy effective theories of gravity admit an ultraviolet completion. The vast research done within this program (see \cite{Palti:2019pca} for a review) has provided evidence that effective theories of quantum gravity satisfy highly non-trivial constraints systematized into a set of interconnected Swampland conjectures. This radically questioned the reliability of many existing non-supersymmetric realizations as consistent solutions in quantum gravity.

While supersymmetric solutions are at least in part protected against quantum corrections and enjoy positive-energy theorems as well as no-force conditions ensuring their stability, the situation for non-supersymmetric solutions is much more delicate, as generically they are subject to all possible sources of instabilities. For Anti de Sitter (AdS) solutions a sharp claim has been made in  \cite{Ooguri:2016pdq,Freivogel:2016qwc}, where it has been conjectured that all such non-supersymmetric solutions supported by flux should be unstable. This was motivated by the well-established Weak Gravity Conjecture \cite{ArkaniHamed:2006dz} and related results such as \cite{Maldacena:1998uz}, leading to the expectation that regardless of the perturbative stability of the solution, a non-perturbative decay channel should always be provided by spontaneous nucleation of membranes discharging the flux that supports the vacuum.

The study of quantum instabilities of metastable spacetimes has a rich history even before and independently of string theory. We mention three among the most relevant examples: the Coleman-de Luccia tunneling \cite{Coleman:1980aw}, the Brown-Teitelboim mechanism \cite{Brown:1988kg} and the nucleation of bubbles of nothing \cite{Witten:1981gj}. Even though the aforementioned decays can be considered as prototypical descriptions of instabilities of string vacua, their explicit construction in concrete stringy setups represents an extremely complicated challenge, both for technical complications in solving the field equations and for the lack of a solid interpretation of the instanton geometries in terms of physical objects in the string spectrum.
Recent studies of non-perturbative instabilities of concrete AdS solutions to higher-dimensional supergravity can be found e.g.\ in~\cite{Apruzzi:2019ecr,Bena:2020xxb,Suh:2020rma,Apruzzi:2021nle}.

In this work, we take a step in this direction by studying the quantum decay of a non-supersymmetric AdS$_4$ vacuum in a concrete compactification  of massive IIA string theory on a six-sphere and other nearly-K\"ahler geometries, originally found in \cite{Lust:2008zd} and further studied in~\cite{Cassani:2009ck,Borghese:2012qm,Guarino:2015vca}. The key feature of our analysis is the implementation of a vacuum decay, embedded in string theory, via expanding D-branes going through a bubble of nothing regime.

Apart from the seminal paper \cite{Witten:1981gj} where bubbles of nothing within the 5d Kaluza-Klein (KK) vacuum were firstly introduced, there are only few examples in the string theory literature regarding this particular decay process. A general obstruction for standard bubble of nothing solutions exists for AdS vacua supported by a Freund-Rubin flux (i.e., a flux filling AdS, or equivalently the full internal manifold) \cite{Young:1984jv}. In \cite{Balasubramanian:2002am} some solutions describing bubbles of nothing in AdS$_5$ were constructed from non-extremal black holes through double analytic continuation and then studied using holography. A class of analytic bubble geometries within AdS spacetimes was worked out in \cite{Dibitetto:2020csn}, but only in the context of gravity plus $\Lambda$. Two important recent works are those of \cite{Ooguri:2017njy}, where such an instanton geometry was constructed within AdS$_5$ vacua of M-theory (with no Freund-Rubin flux) and \cite{GarciaEtxebarria:2020xsr}, where bubbles of nothing were obtained in an Einstein dilaton Gauss-Bonnet model coming from heterotic compactifications.

A particularly interesting scenario is provided in \cite{Horowitz:2007pr}, where the decay of $\text{AdS}_5\times S^5/\mathbb{Z}_k$ vacua of Type IIB is studied. The key idea of this work is to describe the decay process through different classical phases properly linked by dynamical junction conditions. The bubble regime turns out to be the (discharged) bounce geometry interpolating between the AdS vacuum and a phase dominated by a stack of smeared spherical D3 branes that source the whole flux filling the vacuum. It may be worth pointing out that this construction is deeply different from the ones in \cite{Ooguri:2017njy} and \cite{GarciaEtxebarria:2020xsr}. Indeed, in \cite{Horowitz:2007pr}, due to the presence of flux, there is no internal cycle in the geometry shrinking all the way to zero size. As a consequence, the appropriate physical interpretation of this decay process is in line with \cite{Ooguri:2016pdq}, in the sense that it involves the nucleation of spherical D3 branes that start expanding and eventually devour the complete vacuum. 

When it comes to investigating the aforementioned mechanism, a similarly challenging setup turns out to be that of $\AdS_4\times S^6$ vacua in massive IIA string theory. In the present work, we adopt the philosophy of \cite{Horowitz:2007pr} to tackle the issue of their possible non-perturbative instabilities. It is well-known that the compactification of massive IIA over a six-sphere yields non-supersymmetric vacua preserving $\SO(7)$ \cite{Romans:1985tz} as well as ${\rm G}_2$ symmetry \cite{Lust:2008zd,Cassani:2009ck,Borghese:2012qm,Guarino:2015vca}, in addition to a ${\rm G}_2$ invariant supersymmetric vacuum~\cite{Behrndt:2004km}. While in \cite{Borghese:2012qm} the perturbative instability of the $\SO(7)$ invariant vacuum was proven, the ${\rm G}_2$ invariant one turned out to be tachyon free in four-dimensional supergravity. Recently, the latter was also shown to possess a tachyon-free KK spectrum \cite{Guarino:2020flh} and to be protected \cite{Guarino:2020jwv} from the brane-jet decays  that afflict other vacua with stable KK spectrum~\cite{Bena:2020xxb}. These results motivated our interest in the study of non-perturbative decay channels for this vacuum.

In analogy with the case in \cite{Horowitz:2007pr}, the analyzed vacuum is supported by Freund-Rubin flux. As a consequence, the type of bubbling geometry we look for needs to be dressed up by a source and, therefore, will not involve any part of the internal geometry shrinking all the way to zero size. Our main result consists of a fully back-reacted solution in massive IIA supergravity describing the geometry of non-perturbative decay of the ${\rm G}_2$ invariant $\AdS_4$ vacuum. This solution, worked out numerically, interpolates between the vacuum and a smeared D2 brane singularity. The presence of a $\dS_3$ slicing turns out to play a crucial role throughout the whole radial flow. In particular, in the source regime $\dS_3$ is filled by the D2 branes, in an intermediate (approximately neutral) regime it represents the actual expanding bubble, and finally, in the asymptotic region it determines a $\dS_3$ slicing of $\AdS_4$, which crucially dictates the relevant boundary conditions to choose.

An interesting way to study this solution is by looking at the various energy contributions along the spacetime flow. This study manifests the existence of the intermediate regime interpolating between the $\AdS_4$ vacuum and the D2 sources, where the contributions of the Romans' mass, the fluxes, and the internal curvature are sub-dominant with respect to the one coming from the curvature of the bubble. This bubble, though, will turn out to share some properties with a traditional bubble of nothing but it also exhibits some crucial differences. In particular, seen from the viewpoint of a 4d observer living in the vacuum, the bubble is spontaneously created at finite radius and expands, just as a KK bubble would do. Similarly, at the would-be initial location of the bubble a 4d modulus shrinks to zero size. However, in contrast with all existing cases in the literature of bubbles of nothing, such a modulus does not correspond to the physical size of any cycle in the geometry, but rather to the string coupling. Moreover, when at short distances the D2 brane regime kicks in, the profile taken by this shrinking modulus is the one required by D2 branes extending along $\dS_3$ and smeared over the six-dimensional internal manifold.

The non-supersymmetric G$_2$ invariant $\AdS_4$ vacuum of massive IIA on $S^6$, as well as the bubbling solution constructed in this paper, admit a straightforward generalization to the case where the six-sphere is replaced by any other six-dimensional nearly-K\"ahler manifold. Although perturbative stability remains an open question for these other vacuum solutions as their full KK spectrum has not been worked out so far, our analysis indicates that they are non-perturbatively unstable.

\paragraph{Plan of the paper:}
In Section \ref{Summary} we introduce the reader to a non-technical overview of our results, by shedding light on the physical novelties introduced by the bubbles presented in this paper.
In Section \ref{sec:setup} we enter the main body of the paper by introducing the relevant ten-dimensional setup and show how all solutions of this type can be obtained from a lower-dimensional supergravity point of view. After formulating the ansatz, we introduce and discuss the relevant $\AdS_4$ solutions. Next, we start our quest for instabilities. In Section \ref{sec:numerics} we discuss the choice of boundary conditions, present a numerical solution describing a bounce geometry, and analyze its different regimes. In Section \ref{sec:discussion} we conclude by discussing some relevant issues concerning our physical interpretation and possible generalizations of our work. In the appendices we collect our conventions together with additional technical details.

\section{Our results in a nutshell}
\label{Summary}

Starting from the seminal work of \cite{Witten:1981gj}, bubbles of nothing (BoN) arise in the context of KK compactifications of $(d+1)$-dimensional gravitational theories as asymptotically flat and smooth instantons, obtained as the Euclidean continuation of the Schwarzschild black hole. In these solutions, a circle in the compact geometry collapses to zero size at the location of the bubble wall, while a $(d-1)$-dimensional sphere maintains a finite radius. One then turns back to Lorentzian signature by Wick rotating the sphere to de Sitter (dS) space. From the perspective of a lower-dimensional observer, this kind of geometry manifests itself as the nucleation of a BoN with finite radius, which starts expanding and eventually consumes the vacuum. According to such an observer the size of the aforementioned circle is nothing but a shrinking modulus. In this context, it appears to be very natural to try and construct lower-dimensional bubbles where the shrinking modulus is not per se associated with a circle in the geometry.

A special realization of this mechanism can be found within type IIA supergravity. For the sake of our present purpose, we will restrict to effectively 4d bubbles obtained by reductions on a suitable six-manifold $\cM_6$.
Let us start from the associated 11d picture. In this case, the reduction over $\cM_6$ yields an effective 5d gravity description. If $\cM_6$ is Ricci-flat, a $\mathrm{Schw}_5\times \cM_6$ geometry is a valid solution. Its metric reads
\begin{equation}
\dd s_{11}^2 \, = \, -H\,\dd\tau^2 \, + \, \f{\dd\rho^2}{H} \, +\, \rho^2\dd s_{S^3}^2 \, + \,  \dd s_{\cM_6}^2 \ ,
\end{equation}
where 
\begin{equation}
H \, \equiv \, 1-\left(\f{R}{\rho}\right)^2\,,
\end{equation}
 $R$ being the Schwarzschild radius. To obtain the bubble solution of interest we can now perform a double analytic continuation by simultaneously setting
\begin{equation}
	\tau \ \rightarrow \ i\,\psi\,,\qquad S^3 \ \rightarrow \ \mathrm{dS}_3\,,
\end{equation}
where $\psi$ is now a periodic coordinate. Reducing on $S^1_{\psi}$ we end up in the following type IIA background (in string frame)
\begin{equation}\label{BoN_Schw_gauge}
	\begin{aligned}
			\dd s_{10}^2 & =  H^{-1/2}\dd \rho^2 \, +\, \rho^2H^{1/2}\dd s_{\mathrm{dS}_3}^2 \, + \, H^{1/2}\dd s_{\cM_6}^2 \ , \\
		\e^{\Phi} & =  H^{3/4}\,.
	\end{aligned}
\end{equation}
To ensure reality of the fields the range of the radial coordinate has to be restricted to $\rho\,\in\,[R,+\infty)$. Seen from a ten-dimensional perspective, this looks like a BoN where instead of a physical cycle within the geometry, it is the string coupling $\e^{\Phi}$ that shrinks. Since in this paper we will focus on massive type IIA supergravity backgrounds, it is perhaps more appropriate to abandon this eleven-dimensional description in favor of the purely ten-dimensional one. Once in ten dimensions, we should however start by investigating the uniqueness of backgrounds of the form \eqref{BoN_Schw_gauge}.  In fact, a whole one-parameter family of this sort turns out to exist. Its explicit form reads
\begin{equation}\label{BoN_Schw_gen}
	\begin{aligned}
			\dd s_{10}^2 & =  H^{\gamma}\dd\rho^2 \, +\, \rho^2H^{\alpha}\dd s_{\mathrm{dS}_3}^2 \, + \,  H^{\beta}\dd s_{\cM_6}^2 \ , \\
		\e^{\Phi} & =  H^{\delta}\ ,
	\end{aligned}
\end{equation}
where the equations of motion fix 
\begin{equation}
	\beta= \sqrt{\f{1+2\alpha-2\alpha^2}{6}}\,,\qquad \gamma=\alpha-1\,,\qquad \delta=\f{2\alpha-1+\sqrt{6(1+2\alpha-2\alpha^2)}}{4}\,.
\end{equation}
Within this family, the BoN geometry originating from the double analytic continuation and the subsequent reduction of the Schwarzschild solution corresponds to $\alpha=\f{1}{2}$.

\begin{figure}[!htb]
	\centering
	\includegraphics[width=0.6\textwidth]{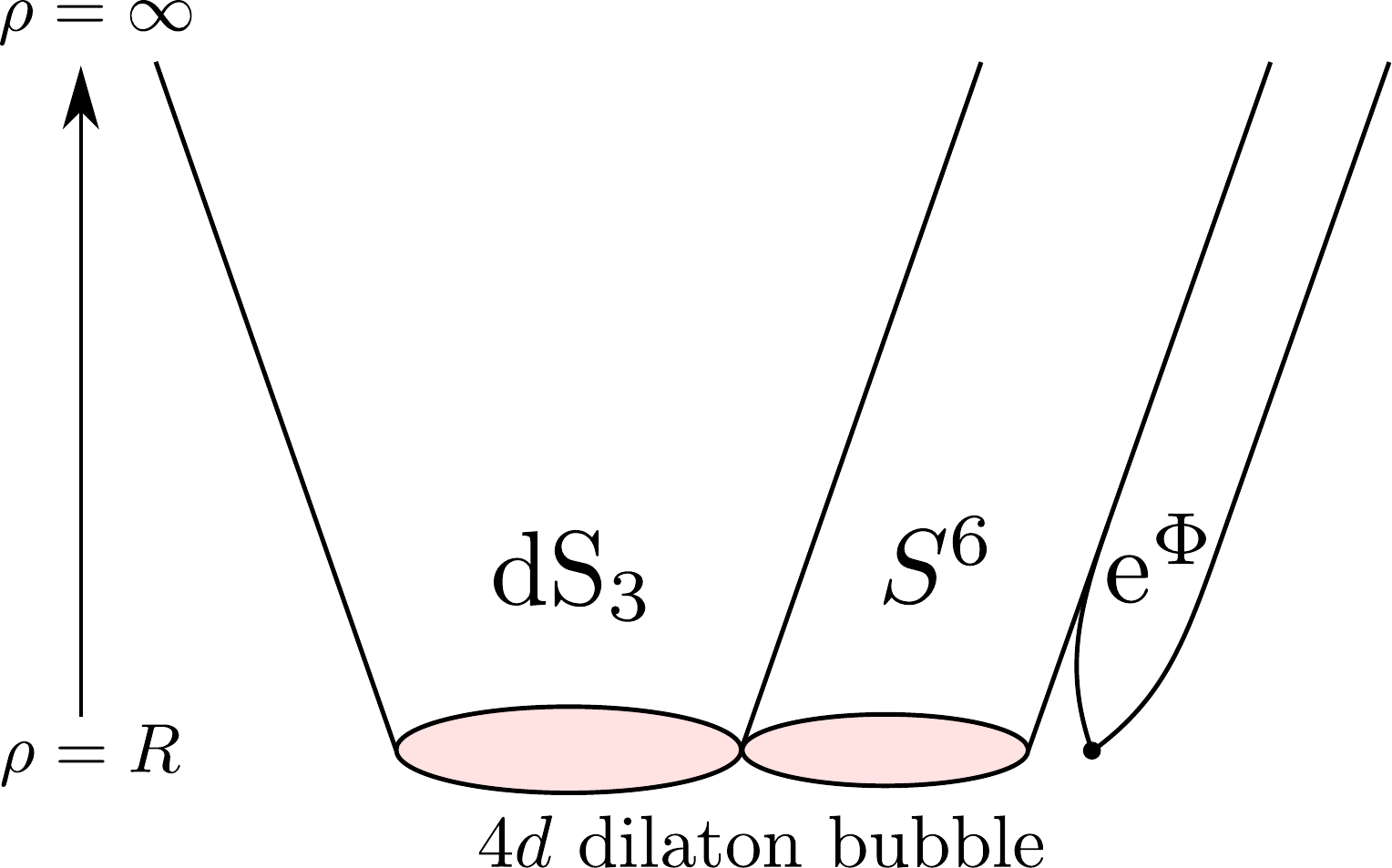}
	\caption{\it An artistic impression of a "dilaton bubble". According to a 4d observer in its Einstein frame, a BoN is created at finite radius $R$ and starts expanding out towards infinity. At the bubble wall the string coupling shrinks to zero, while the overall size of the internal manifold stays stable.}
	\label{fig:Dilaton_bubble}
\end{figure}

Taking a closer look at the physical properties of the solution, the case $\alpha=\f{1}{2}$ is singled out as the only choice such that $\alpha=\beta$ and the geometry admits a frame in which both $\dS_3$ and $\cM_6$ have a finite size at $\rho=R$.\footnote{By ``choice of frame'' we mean here a Weyl transformation where the Weyl factor is given by a chosen power of $\e^\Phi$. The condition $\alpha=\beta$ is also achieved by taking $\alpha = -1/4$, however in this case the dilaton is constant (as $\delta=0$) and there is no frame such that dS$_3$ and $\cM_6$ have a finite size at $\rho=R$.} This frame is obtained by multiplying the metric in \eqref{BoN_Schw_gen} by $\e^{-\f{2}{3}\Phi}$ and corresponds to both the eleven and four-dimensional Einstein frame. In other words, an observer living in the 4d vacuum experiences a BoN with finite radius $R$ which is suddenly created at the center of space and immediately starts to expand, eventually consuming the complete vacuum. A conceptual picture of this situation is depicted in Figure \ref{fig:Dilaton_bubble}. On the other hand, from the ten-dimensional perspective this bubble has a different origin than the traditional BoN solutions, as in our case the string coupling rather than an internal cycle shrinks to zero size. However, as far as the four-dimensional observer is concerned the resulting effective description is identical, thus unifying KK and string coupling instabilities. 

The scope of this paper is to investigate the non-perturbative (in)stability of certain $\AdS_4$ vacua obtained from compactifying massive IIA supergravity on a nearly-K\"ahler manifold $\cM_6$. The main result is the existence of a complete charged bubble membrane that expands and eats up the analyzed vacuum. The above dilaton bubble solution will be a fundamental building block in the whole construction. Nevertheless, in the situation at hand such a bubble will only be a good approximate description in an intermediate regime. 
\begin{figure}[!htb]
	\centering
	\includegraphics[width=0.7\textwidth]{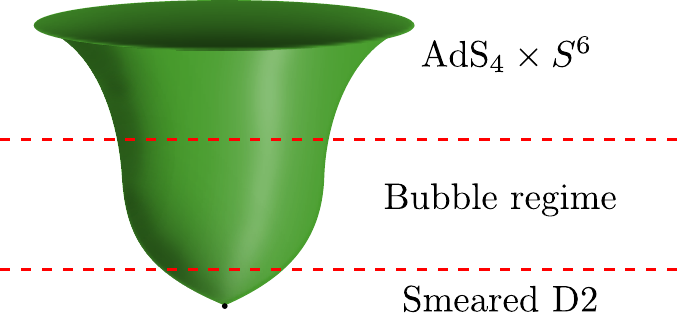}
	\caption{\it A conceptual picture of the radial evolution of the bubble geometry. The "dilaton bubble" is a good intermediate description. In the large distance limit it connects to the vacuum phase. Moreover, the bubble regime cannot persist indefinitely at small distances and it ultimately flows to a (smeared) D2 brane source.}
	\label{fig:Dbubble}
\end{figure} 
This will be due to the fact that several extra ingredients necessary to construct the vacuum crucially imply some new emergent features. In particular, the presence of $\AdS_4$ in the asymptotic region requires a modified large distance behavior, while on the other hand, a non-zero Freund-Rubin flux determines a modified short distance behavior induced by the presence of smeared D2 brane sources. Finally, two extra complications are represented by a non-zero Romans' mass as well as by the departure of $\cM_6$ from Ricci-flatness.\footnote{Indeed, starting from the next section,  $\cM_6$ will be firstly taken to be a generic nearly-K\"ahler manifold, and secondly further specified to a six-sphere.} However, although they are both crucial for the very existence of the asymptotic vacuum, in the intermediate regime these ingredients will turn out to only yield subleading modifications of the 
bubble geometry. An intuitive representation of the different phases of the resulting geometry is given in Figure \ref{fig:Dbubble}.

\section{The supergravity setup}
\label{sec:setup}

We are interested in solutions of massive type IIA supergravity representing bounce geometries which asymptotically approach AdS$_4\times \cM_6$, where $\mathcal{M}_6$ is $S^6$, or more generally a compact nearly-K\"ahler manifold. As in the original bubble of nothing introduced in \cite{Witten:1981gj}, the four-dimensional geometry should represent a domain wall solution with curved three-dimensional de Sitter slices. 

\subsection{Ten-dimensional ansatz}

Our aim thus consists in constructing curved domain wall geometries preserving the symmetry of $\dS_3$ as well as the nearly-K\"ahler (NK) structure on $\mathcal{M}_6$. For the particular case where $\mathcal{M}_6 = S^6$, the latter statement is equivalent to declaring that we preserve the G$_2$ symmetry acting on $S^6\simeq {\rm G}_2/{\rm SU}(3)$. The most general ten-dimensional ansatz respecting these symmetries is given by\footnote{In principle one could contemplate adding a term of the form $\f{1}{2g^3}\dd \xi(r) \wedge \Re\Omega$ to $F_4$ and a corresponding term to $F_6$. Such a term is consistent with the Bianchi identity for $F_4$, however the Bianchi identity for $F_6$ forces $\dd\xi=0$. Hence, this term is pure gauge and does not appear in the field strengths.}
\begin{equation}\label{10ansatz} 
	\begin{aligned}
	\ds_{10}^2 \,&=\, \e^{2V(r)}\left( \dd r^2 + L^2\e^{2A(r)} \dd s^2_{\dS_3}  + g^{-2}\ds_{\cM_6}^2 \right)\,,\\[1mm]
	\e^{\Phi} \,&=\, \e^{\phi(r)}\,,\\
	B_2 \,&=\, g^{-2}\, b(r) J	\,,\\
	F_0 \,&=\, m\,,\\ 
	F_2 \,&=\, F_0 B_2\,,\\ 
	F_4 \,&=\, f_6(r) \star \dvol_{\cM_6} + f_{41}(r)J\wedge J 
	+ f_{42}(r)\,\dd r\wedge \Im\Omega\,,
	\end{aligned}
\end{equation}
where $m\,,g$ and $L$ are constant parameters and a prime denotes a derivative with respect to the radial coordinate $r$. While $m$ is simply the Romans mass, $g$ provides an overall scale for the internal manifold and will also control the internal $F_6$ flux. The metric $\dd s^2_{10}$ is taken in the string frame, $\ds^2_{\dS_3}$ is the unit metric on $\dS_3$ and $\ds_{\cM_6}^2$ is the NK metric on $\cM_6$, normalized such that the Ricci tensor equals $5$ times the metric\footnote{When $\cM_6 = S^6$, $\ds^2_{\cM_6}$ is the unit metric on the round sphere. A convenient way to parameterize the metric on dS$_3$ is $$\ds_{\dS_3}^2 = -\dd t^2 + \cosh^2 t \,(\dd \theta^2 + \sin^2\theta \dd\phi^2)\,.$$ This makes the Wick rotation to $S^3$ straightforward, which will be useful for our purposes.}. Note that the function multiplying $\dd r^2$ is a gauge choice: we have chosen to take it proportional to the conformal factor of the NK metric on $\mathcal{M}_6$ as this will be the natural choice to describe the space transverse to D2 brane sources. The forms $J$ and $\Omega$ are respectively a real two-form and complex three-form defining the NK structure on $\cM_6$ (for more details see Appendix~\ref{app:S6}). Denoting by $\dvol_{\dS_3}$ and $\dvol_{\cM_6} = \f{1}{6}J\wedge J\wedge J$ the volume forms on dS$_3$ and $\mathcal{M}_6$, respectively, our ten-dimensional volume form reads
\begin{equation}
\dvol_{10} \,=\, L^3 g^{-6}\e^{3A+10V}\,{\dvol}_{\dS_3} \wedge \dd r \wedge \dvol_{\cM_6}\,,
\end{equation}
which also fixes the orientation. The NS-NS field strength is given by
\begin{equation}
H_3 = \dd B_2 =   g^{-2}\,b'\,\dd r\wedge J+ 3\, g^{-2} b \,\Re\,\Omega\,.
\end{equation}
The dual R-R field strengths are given by
\begin{align}
	F_6 \,&=\,\star F_4 \,=\, -f_6\,{\rm vol}_{\cM_6} + 2L^3g^2 f_{41}\,\e^{3A+2V}{\dvol}_{\dS_3} \wedge \dd r\wedge  J - L^3 f_{42}\, \e^{3A+2V}{\rm vol}_{\dS_3} \wedge \Re\,\Omega\, \,,\nn\\
	F_8 \,&=\,-\star F_2 \,=\,  - \f{mb L^3}{2g^4}\,\e^{3A+6V}\,{\rm vol}_{\dS_3} \wedge \dd r\wedge  J\wedge J\,,\\
	F_{10}\,&=\,\star F_0 \,=\, m\, {\dvol}_{10}\,.\nn
\end{align}
In order to satisfy the Bianchi identity $\dd F_4 = H_3 \wedge F_2$ we have to impose
\begin{align}\label{bianchifluxes}
	f_6 &= -\f{1}{g^2}\big( m b^3 + 6g b \zeta + 5g \big)\,,\nonumber\\
	f_{41} &= \f{1}{2g^4} \big( m b^2 + 2g\zeta \big)\,,\\ 
	f_{42} &= \f{1}{2g^3} \,\zeta^\prime\,,\nonumber
\end{align}
while the Bianchi identity for $F_6$ --- that is, the $F_4$ equation of motion --- gives a second-order equation for $\zeta$,
\begin{equation}
	\zeta^{\prime\prime} + \left( 3A^\prime + 2 V^\prime \right)\zeta^\prime - 6g\left( (mb^2+2g\zeta) + \e^{-4V}b(5g+mb^3+6gb\zeta)\right) \,=\,0 \,.
\end{equation}
The equations for the other R-R fields are solved automatically by imposing this ansatz. Additionally, one can straightforwardly evaluate the equations of motion for the NS-NS fields on our ansatz. However, in order to present our final system of equations in a compact elegant way, it is more convenient to exploit a consistent truncation to four dimensions, as we describe next.

\subsection{Effective action}

The ansatz described above is contained in a consistent truncation of massive IIA supergravity on any NK manifold $\mathcal{M}_6$ \cite{Kashani-Poor:2007nby,Cassani:2009ck}.\footnote{We will use the notation of \cite[sec. 4.2]{Guarino:2015vca}, except for the scalar $\ti\xi$, which we denote here by $\zeta$.} The result of this truncation yields a four-dimensional $\mathcal{N}=2$ supergravity coupled to one vector multiplet and one hypermultiplet, with an Abelian, dyonic gauging of certain hyperscalar isometries. The  bosonic fields in the four-dimensional theory are the metric $\dd s_4^2$, five scalars ($U$, $\phi$, $b$, $\xi$, $\zeta$), two vectors $a_1$, $b_1$ and a two-form $b_2$. The gauging is controlled by the two parameters $g$ and $m$. The symmetries of the solutions we are interested in lead us to set $a_1=b_1=b_2=0$ and $\xi$ equal to an arbitrary constant, consistently with their equations of motion.\footnote{The value of the constant does not matter since  $\xi$ only enters the Lagrangian through its derivative.} The action for this subtruncation is given by 
\begin{equation}\label{4daction}
\begin{aligned}
		S_4 =& \,\f{1}{2\kappa_4^2}\int_{\cM_4}\!\! \dvol_4 \left(\cR - 24 (\partial U)^2 - \f12\,(\partial \phi)^2 - \f32\,\e^{-\phi-4U} (\partial b)^2 - \f12\, \e^{\f{\phi}{2}-6U}(\partial \zeta)^2 - \mathcal{V} \right) \\[1mm]
	 	&+ \f{1}{\kappa_4^2}\int_{\partial\cM_4} {\dvol}_{\partial\cM_4}\,\cK \,,
	\end{aligned}
\end{equation}
where the scalar potential $\mathcal{V}$ is given by
\begin{equation}\label{NKpot}
	\begin{aligned}
		\mathcal{V}\,=\, \f12\e^{-\f{\phi}{2}-18U}(\,5g\, +&\, mb^3 + 6g b\zeta\,)^2 + \f32\e^{\f{\phi}{2}-14U}(\,mb^2+2g\zeta\,)^2 \\ &-30g^2\e^{-8U}+18g^2 \e^{-\phi-12U} b^2 + \f32 m^2 \e^{\f32\phi-10U} b^2 + \f12 m^2 \e^{\f52\phi-6U}\,.
	\end{aligned}
\end{equation}
The boundary term is the Gibbons-Hawking-York term ensuring a good variational problem for the four-dimensional metric in an asymptotically locally AdS spacetime. Here ${\dvol}_{\partial\cM_4}$ and $\mathcal{K}$ are the volume form and the trace of the extrinsic curvature on the conformal boundary $\partial\cM_4$.

The variables appearing here match our ten-dimensional ansatz \eqref{10ansatz}, \eqref{bianchifluxes} upon giving just a radial dependence to all scalar fields, identifying the conformal factor $V$ of the ten-dimensional metric as
\begin{equation}
 V\, =\, U + \f{\phi}{4}\,,
\end{equation} 
and choosing the four-dimensional metric as
\begin{equation}\label{4dmetric}
\ds^2_4 \,=\, \e^{8U(r)}\left(\dd r^2 + L^2\,\e^{2A(r)}\, \ds^2_{\dS_3} \right)\,,
\end{equation}
where again $\ds^2_{ \dS_3}$ is the unit metric on  $\dS_3$. The four-dimensional Ricci scalar then reads 

\begin{equation}
\cR \,=\, 6\, \e^{-8U} \left(\f{1}{L^2}\,\e^{-2A} - 2\, (A'+2U')(A'+4U') -  A'' -4U'' \right)\,,
\end{equation}
the trace of the extrinsic curvature is
\begin{equation}
\mathcal{K} = 3\, \e^{-4U}\left(A'+4U'\right)\,,
\end{equation}
while the 4d volume form is given by
\begin{equation}
{\dvol}_4 \,=\, L^3\,\e^{3A+16U}\,{\dvol}_{\dS_3}\wedge \dd r\,.
\end{equation}
Plugging this in~\eqref{4daction}, we arrive at the one-dimensional effective action 
\begin{align}\label{1dAction}
	S_{\rm 1}&=  \f{L^3\vol_3}{2\kappa_4^2} \!\int\! \dd r\left[ \e^{3A+8U}\!\left(\! 6 (A'+4U')^2 - 24 (U')^2 - \f12(\phi')^2 - \f32\e^{-\phi-4U}(b')^2 -\f12\, \e^{\f{\phi}{2}-6U}(\zeta')^2 \!\right)\right.\nn\\[1mm]
	&\hspace{2.8cm}\left. 
	 - \,\e^{3A+16U}\mathcal{V}  + \f{6}{L^2}\,\e^{A+8U} \right]   \,,
\end{align}
where $\vol_3 = \int \dvol_3$ is the volume of the three-dimensional geometry.\footnote{Of course $\vol_3$ is divergent as de Sitter space is non-compact. In order to give it a finite value we can analytically continue $\dS_3$ to $S^3$ and use $\int_{S^3}{ \dvol}_{S^3}\,=\,2\pi^2$.}
 As usual when considering a domain-wall ansatz like \eqref{4dmetric}, the second-order equations generated by varying the one-dimensional action should be supplemented with a constraint, coming from the $rr$ component of the Einstein equation. This gives a first-integral of the second-order equations, that has the physical meaning of setting to zero the radial Hamiltonian of the system. In our case this Hamiltonian constraint reads
\begin{equation}\label{HamConstr}
\begin{aligned}
&\e^{3A+8U}\left( 6 (A'+4U')^2 - 24 (U')^2 - \f12(\phi')^2 - \f32\e^{-\phi-4U}(b')^2 -\f12\, \e^{\f{\phi}{2}-6U}(\zeta')^2 \right)\\ 
&+\, \e^{3A+16U}\mathcal{V}  - \f{6}{L^2}\,\e^{A+8U} \,=\,0\,.
\end{aligned}
\end{equation}
We have checked that the equations of motion generated by the action \eqref{1dAction} together with the first integral \eqref{HamConstr}, are equivalent to the equations obtained by plugging \eqref{10ansatz}, \eqref{bianchifluxes} into the massive IIA supergravity equations.

In the following we will continue our analysis by choosing the values of the parameters as
\begin{equation}\label{choiceparameters}
m=1\,,\quad g=1/2\,.
\end{equation}
With this choice we do not lose any generality since, as shown in Appendix~\ref{AppRescaling}, any other choice of the parameters can easily be related to the one above by suitable rescalings.

\subsection{The set of AdS$_4$ vacua}\label{sec:AdSvacua}

The model above contains $\AdS_4$ vacuum solutions, which uplift to $\AdS_4\times \mathcal{M}_6$ solutions of massive IIA supergravity. These are obtained by requiring that the scalar potential is extremized and setting the $\dS_3$ warp factor to\footnote{We denote by a subscript ``$0$'' the scalar vev's at the extrema.}
\begin{equation}
\e^{A} \,=\, \e^{-4U_0}\sinh \left(\f{r}{{\e^{-4U_0} L}}\right)\,.
\end{equation}
In this way the 4d metric \eqref{4dmetric} describes an AdS$_4$ space foliated by $\dS_3$ leaves, with characteristic length $L$ and cosmological constant $\Lambda=\f{1}{2}\mathcal{V} = -\f{3}{L^2}$.
 One finds three such extrema~\cite{Cassani:2009ck} for which the values of the scalars and the AdS length are reported in Table~\ref{tab:AdSvac}. One vacuum preserves $\cN=1$ supersymmetry while the remaining two are non-supersymmetric.\footnote{From a 10d perspective, the first non-supersymmetric vacuum in Table~\ref{tab:AdSvac} was constructed within massive type IIA supergravity in~\cite{Romans:1985tz},  the second was first obtained in~\cite{Lust:2008zd}, while the supersymmetric one was first found in~\cite{Behrndt:2004km}.}
\begin{table}[!htb]
	\centering
	\begin{tabular}{ccccccc}
		SUSY  &  $G_{\rm res}$ & $\e^{2\phi_0}$& $\e^{8 U_0}$& $\zeta_0$& $ b_0$ & $L^2$\\[1mm]
		\hline
		$\cN = 0$ & $\SO(7)$ & $\f{5^{5/6}}{2^{5/3}}$& $\f{2^{1/3}}{5^{1/6}}$& $0$ & $0$ &$\f{2^{10/3}}{5^{7/6}}\Tstrut$\\[2mm]
		$\cN = 0$ & ${\rm G}_2$ & $\f{\sqrt{3}}{2}$& $\f{3\sqrt{3}}{8}$& $-1$ & $\f12$ &$\f{3\sqrt{3}}{4}\Tstrut$\\[2mm]
		$\cN = 1$ & ${\rm G}_2$ & $\f{\sqrt{15}}{4}$& $\f{15\sqrt{15}}{64}$& $\f12$ & $-\f14$ & $\f{25\sqrt{15}}{64}\Tstrut$
	\end{tabular}
	\caption{The values of the scalars and AdS$_4$ length at the critical points of the potential. 
	$G_{\rm res}$ denotes the residual symmetry of the ten-dimensional solution for the case where $\mathcal{M}_6=S^6$.}
	\label{tab:AdSvac}
\end{table}

For the sake of understanding the correct perturbative regime in which the supergravity solution is reliable, it is convenient to
reinstate for a moment the parameters $m$ and $g$. By doing so, we have that $\e^{\phi} \sim (\f{g}{m})^{5/6}$ and $\e^{U}/g\sim \f{m^{1/24}}{g^{25/24}}$ for the string coupling constant and the internal length scale (see Appendix~\ref{AppRescaling}). We can trust the 10d supergravity regime when the former is small and the latter is large. This is achieved by taking $m\gg g$ and, for instance, $g = \mathcal{O}(1)$. Although for the sake of a cleaner presentation we work with the $\mathcal{O}(1)$ choice \eqref{choiceparameters}, the choice ensuring validity of the supergravity regime can be straightforwardly related to this one via the rescalings discussed in Appendix~\ref{AppRescaling}.

All three vacua presented above are perturbatively stable against fluctuations of the scalars in the action \eqref{4daction}~\cite{Cassani:2009ck}. Of course, this represents a very limited sector of all KK modes of massive IIA supergravity. Studying the full KK spectrum is in general extremely complicated, however some interesting results have been obtained by focusing on the case where the internal space is a sphere, $\mathcal{M}_6 = S^6$. For the $\SO(7)$ vacuum one can quickly confirm the expectation that a non-supersymmetric vacuum is unstable as this solution is already unstable against perturbations within the consistent truncation leading to four-dimensional maximal $\ISO(7)$ supergravity \cite{Guarino:2015vca}. The non-supersymmetric ${\rm G}_2$ invariant vacuum on the other hand is perturbatively stable against all 70 independent scalar fluctuations in the $\ISO(7)$ theory and moreover it has recently been shown to remain perturbatively stable even when one considers the full KK spectrum of massive type IIA supergravity \cite{Guarino:2020flh}. In addition, this vacuum can be shown \cite{Guarino:2020jwv} to be stable against possible brane-jet decays of the type discussed in~\cite{Bena:2020xxb}. This therefore leaves us with the question whether this solution is ultimately stable and violates the AdS Swampland conjecture or if there exists another non-perturbative decay channel. In the remainder of this paper we will deal with this question and construct a solution that deforms the ${\rm G}_2$ invariant $\AdS_4$ vacuum, as well as the similar vacua obtained by replacing $S^6$ with a nearly-K\"ahler $\cM_6$, while obeying the same asymptotic boundary conditions.

\section{The decay channel: explicit solutions}
\label{sec:numerics}

We start our search for a bounce geometry illustrating the non-perturbative decay of the perturbatively stable $\AdS_4$ vacuum with residual G$_2$ symmetry, which we introduced above. Since the asymptotic $\AdS_4$ solution does not preserve supersymmetry, there is little hope to find an analytic expression for the bounce geometry. However, we can still solve the equations of motion by resorting to numerical methods. It may be worth noticing that a successful numerical integration necessarily goes through the problem of understanding how to assign suitable boundary conditions.
Once an explicit example of such a numerical flow is given, its physical features turn out to crucially depend on its boundary behavior at the end of the radial flow.
Namely, if e.g.\ the 10d geometry is smoothly capped off, what we have is a standard BoN, while the case of a singularity might signal the presence of a source.

When first considering the existence of a KK bubble where the internal space $\cM_6$ shrinks to zero size as a potential instability, a simple scaling argument analogous to \cite{Young:1984jv} shows its inconsistency. This is due to the presence of a Freund-Rubin type flux on $S^6$, which induces  divergent terms in the field equations in the limit where internal space shrinks.  Hence, we must include a source at small distances in order to modify the inconsistency in the field equations. We point out that this fact is compatible with a more general topological argument. In fact smooth bubble geometries can be represented as the warped product of $\dS_3$ with a 7d manifold $\cM_7$ such that $\partial \cM_7=\cM_6$. A combination of Stokes' theorem and Bianchi identities would then imply the vanishing of $F_6$ through the 6-sphere, which is obviously inconsistent with our setup. It is again the inclusion of a source carrying the same $F_6$ flux as the vacuum that restores charge conservation.

This indicates that any pure bubble geometry --- both of the KK type as well as the dilaton bubble geometries introduced in this work --- can only be an approximate solution in some intermediate regime, while a source regime is expected to take over at small distances, as already shown in Figure \ref{fig:Dbubble}. 
The source in our case has to be a (smeared) D2 brane, since the relevant flux is given by $F_6$. For this reason, our situation is very similar to that of~\cite{Horowitz:2007pr}, which we will take as a guide reference for our analysis.

\subsection{Linearized boundary analysis}
\label{sec:asymptotics}

Our numerical approach towards constructing the bubbling solution will consist of integrating the equations of motion starting from the conformal boundary. The initial conditions are thus defined by specifying how the different fields are deformed away from the $\AdS_4$ vacuum in the asymptotic  near-boundary region. We can discuss this by recalling some  basic features of asymptotically AdS solutions. 
By a suitable change of radial coordinate $z=z(r)$, the metric \eqref{4dmetric} can be cast in the Fefferman--Graham form 
\begin{equation}
\dd s^2_4 \,=\, \f{L^2}{z^2}\, \left( \dd z^2  + \gamma(z)\,\ds^2_{\dS_3}\right)\,,
\end{equation}
where $\gamma(z)$ has the expansion $\gamma(z) \,=\, \gamma_{0} + z^2 \gamma_{2} + z^3 \gamma_{3} + \ldots\,$. Note that in the new radial coordinate the conformal boundary is found at $z=0$. A scalar field $\Phi$ of mass $M$ has a non-normalizable and a normalizable mode; the asymptotic behavior of these modes is determined by a positive solution $\Delta$ to the indicial equation\footnote{In a holographic context, $\Delta$ corresponds to the conformal dimension of the dual CFT operator.}
\begin{equation}
\Delta(\Delta-3) = M^2 L^2\,.
\end{equation}
 Since we are interested in solutions that match the $\AdS_4$ vacuum asymptotically, we demand that only the normalizable modes are activated. Then the asymptotic expansion of the scalar reads
\begin{equation}\label{scalFG}
\Phi(z) = \Phi_0 + z^\Delta \,\Phi_1 + \ldots \,,
\end{equation}
where $\Phi_0$ is the constant AdS value and the dots denote subleading terms. The initial conditions for our numerical analysis will be specified by choosing $\Phi_1$, the rest of the solution being then entirely fixed in terms of this.

We can compute the scalar masses by diagonalizing the mass matrix obtained from the action~\eqref{4daction}. For the second vacuum in Table~\ref{tab:AdSvac}, i.e.\ the non-supersymmetric ${\rm G}_2$ invariant vacuum, the mass eigenvalues are
\begin{equation}
M^2 L^2 \,=\  6\,,\ 6\,,\ 20\,,\ 20\,,
\end{equation}
corresponding to
\begin{equation}
\Delta\,=\quad  \f{3+\sqrt{33}}{2}\,,\quad  \f{3+\sqrt{33}}{2}\,,\quad \f{3+\sqrt{89}}{2}\,,\quad \f{3+\sqrt{89}}{2}\,.
\end{equation}
Note that these are pairwise equal, namely the associated eigenspaces are two-dimensional; this will play a role in our discussion later. 
Using the linear transformation relating the mass eigenstates and the scalar fluctuations over the vacuum, we find that the asymptotic behavior of our scalar fields as $z\to 0$ is given by
\begin{equation}\label{asymptotic_expr_scalars}
\begin{aligned}
U \,&=\, U_0 \,+\,  \f{4c_1-3c_2}{28}\, z^{\f{3+\sqrt{33}}{2}} + c_4\,z^{\f{3+\sqrt{89}}{2}} + \ldots \,,\\[2mm]
b \,&=\, b_0 \,+\,  \f{3 c_2-4c_1}{7}\, z^{\f{3+\sqrt{33}}{2}} + \f{7c_3 + 36 c_4}{12}\,z^{\f{3+\sqrt{89}}{2}} + \ldots\,,\\[2mm]
\phi \,&=\, \phi_0 \,+\, c_2\, z^{\f{3+\sqrt{33}}{2}} - c_3\,z^{\f{3+\sqrt{89}}{2}} + \ldots \,,\\[2mm]
\zeta\,&=\, \zeta_0 \,+\,  c_1\, z^{\f{3+\sqrt{33}}{2}} + c_3\,z^{\f{3+\sqrt{89}}{2}} + \ldots\,,
\end{aligned}
\end{equation}
where $c_1,c_2,c_3,c_4$ are free coefficients which completely determine the subleading terms. There is no additional free constant in the near-boundary expansion of the metric function $A$. The expression \eqref{asymptotic_expr_scalars} can be transformed back to the original radial variable $r$ by using the asymptotic form of the transformation defining the Fefferman--Graham radial coordinate,
\begin{equation}
z = \e^{- r/\ell} + \ldots\,,\qquad \text{with}\ \ \ell \equiv \e^{-4U_0}L=\sqrt{2}\,.
\end{equation}
As a cross-check, we have directly verified that \eqref{asymptotic_expr_scalars} solve the linearized equations of motion near the conformal boundary. We will use the asymptotic solution~\eqref{asymptotic_expr_scalars} as initial conditions for our numerical analysis, after choosing a value for the free coefficients. Such a choice of boundary conditions as input for the numerical integration then guarantees the correct AdS asymptotics of the flow.

\subsection{Different regimes for the 10d geometry}
 Analogous to the situation in \cite{Horowitz:2007pr}, our numerical solution will turn out to interpolate among three regimes: the asymptotic $\AdS$ regime, the intermediate bubble regime and finally the source regime, matching the picture anticipated in Figure~\ref{fig:Dbubble}.

\textbf{AdS regime:} As explained above, in the numerical analysis we carefully choose the boundary data at large $r$ to only perturb our $\AdS_4$ vacuum with normalizable modes. This choice therefore guarantees that the asymptotic solution will not be affected by the perturbation. This is indeed reflected in our numerical analysis by the fact that for large values of the radial coordinate we always approach the correct vacuum. 

\textbf{Bubble regime:} In analogy with \cite{Horowitz:2007pr}, our solutions exhibit an intermediate regime  corresponding to a bubble geometry. Such a background consists of an asymptotically flat geometry of the form \eqref{BoN_Schw_gauge}. However, in contrast with this reference, in our case the bubble is described by a 4d dilaton bubble where the string coupling shrinks to zero size at the location of the bubble wall. As already discussed earlier, such a bubble will not be a solution to the full dynamical system we are studying, but rather an approximate solution valid in a regime where the fluxes as well as the curvature of internal space can be neglected.

\textbf{Source regime:} As argued before, in the presence of a Romans' mass and flux along $\cM_6$, the bubble geometry described above is in fact inconsistent as a solution to the full system of equations. This implies that the bubble regime cannot persist indefinitely and must be modified at small values of the radial coordinate $r$ to include brane sources.

In our case the appropriate object to support the bubble is given by D2 brane sources carrying the same flux as the AdS vacuum and smeared over the entire internal space. This geometry is described by the following massive type IIA background,
\begin{equation}\label{smearedD2}
	\begin{aligned}
			\dd s_{10}^2 & = H_{\mathrm{D}2}^{-1/2}\dd s_{\mathrm{dS}_3}^2\,+\, H_{\mathrm{D}2}^{1/2}\left(\dd r^2\,+\, g^{-2}\dd s_{\cM_6}^2\right) \ ,\\
		\e^{\Phi} & = H_{\mathrm{D}2}^{1/4} \ ,\\
		C_{3} & = \left(H_{\mathrm{D}2}^{-1}-1\right) \, \mathrm{vol}_{\mathrm{dS}_3}\ ,
	\end{aligned}
\end{equation}
where, due to the smearing of the D2's over $\cM_6$, the explicit form of the harmonic function is linear in $r$,
\begin{equation}
	H_{\mathrm{D}2} = Q_{\mathrm{D}2}(r-r_{\mathrm{D}2})\,,
\end{equation}
where $Q_{\mathrm{D}2}=5g^5$ gives the flux 
\begin{equation}
\int F_6 = \int \star\, \dd C_3 = \frac{Q_{\rm D2}}{g^6}\int{\rm vol}_{\cM_6} = \frac{5}{g}\int{\rm vol}_{\cM_6}
\end{equation} 
that matches the Page charge $\int F_6^{\rm Page}$ of the AdS$_4\times \cM_6$ vacuum, while $r_{\mathrm{D}2}$ is the position of the source, which will be fixed by the dynamics. Just as for the BoN geometry, the curved, smeared D2 brane described in equations \eqref{smearedD2} is only an approximate solution valid in a regime where the curvatures of the $\dS_3$ and $\cM_6$ can be neglected. However, similar as in \cite{Bobev:2018ugk} this background can be made into a full curved brane solution by turning on additional axionic scalars as can be seen in the numerical solution. 

It may be worth stressing that when $\cM_6= S^6$ such a smeared source, contrary to its localized counterpart, cannot be interpreted as a consistent solution within the G$_2$ invariant sector of ISO$(7)$ gauged supergravity. For this reason one has to consider the NK truncation, which describes the full set of G$_2$ invariant fields, in order to find the instability discussed here. Consistently with this result, in \cite{Dibitetto:2021ltm} a positive energy theorem for this vacuum has been proven within the G$_2$ invariant sector of the ISO$(7)$ theory.

\textbf{Gluing the three regimes:} Having introduced the three regimes, one can see that they can be consistently glued together. To do so it is convenient to re-express the BoN metric \eqref{BoN_Schw_gauge} in an appropriate gauge for the radial coordinate to be compatible with the D2 brane metric in \eqref{smearedD2}. Namely, we make a change of coordinate $\rho = \rho(r)$ satisfying
\begin{equation}\label{D2_gauge}
H^{1/2} = H^{-1/2}\left(\f{\dd\rho}{\dd r}\right)^2 \,,
\end{equation}
so that the radial and $\cM_6$ components of the metric \eqref{BoN_Schw_gauge} carry the same radial dependence, as in \eqref{smearedD2}.\footnote{This is also the choice we made in the metric \eqref{10ansatz}.}
This condition is solved by
\begin{equation}
	\rho^2 = R^2\,+\,(r-r_{\rm B})^2\,,
\end{equation}
where $r_{\rm B}$ is an integration constant representing the radial position of the BoN wall in the solution \eqref{BoN_Schw_gauge}. Summarizing, for this choice of the radial coordinate, the metric functions as well as the dilaton are given by the following expressions in the various regimes:
\begin{equation}\nn
\begin{array}{l|l|l}
	\multicolumn{1}{c}{\textbf{Source regime}} & \multicolumn{1}{c}{\textbf{Bubble regime}} & \multicolumn{1}{c}{\textbf{AdS regime}} \\[1mm]
	\hline
	\e^{U+A+\f{\phi}{4}} = \e^{U_{\rm S}} H_{\rm D2}^{-1/4} & \e^{U+A+\f{\phi}{4}} = \f{ \e^{U_{\rm B}}(r-r_{\rm B})^{1/2}}{L}\left(R^2+(r-r_{\rm B})^2\right)^{1/4} & \e^{U+A+\f{\phi}{4}} = \frac{2}{\sqrt3}\sinh \f{r}{\sqrt2}\Tstrut\\[4mm]
	\e^{U+\f{\phi}{4}} = \e^{V_{\rm S}} H_{\rm D2}^{1/4} & \e^{U+\f{\phi}{4}} = \e^{U_{\rm B}} \f{(r-r_{\rm B})^{1/2}}{\left(R^2+(r-r_{\rm B})^2\right)^{1/4}} & \e^{U+\f{\phi}{4}} = (\frac{3}{4})^{1/4} \\[4mm]
	\e^\phi = \e^{5V_{\rm S}} H_{\rm D2}^{1/4} & \e^{\phi} =  \e^{\Phi_{\rm B}}\f{(r-r_{\rm B})^{3/2}}{\left(R^2+(r-r_{\rm B})^2\right)^{3/4}} & \e^\phi = (\frac{3}{4})^{1/4} 
\end{array}
\end{equation}
The AdS regime is the one discussed in Sec.~\ref{sec:AdSvacua}; the constants $U_{\rm B}$, $\Phi_{\rm B}$, $U_{\rm S}$ and $V_{\rm S}$ are inserted in order to be more general in the matching with the bubble and source regimes and still solve the equations of motion at leading order. The proper gluing conditions will be satisfied for the explicit solution by a suitable choice of these constants.

\subsection{Explicit numerical solutions}

We are now ready to use the expressions~\eqref{asymptotic_expr_scalars} as initial conditions for our numerical analysis, after choosing values for the four free coefficients. Indeed, by construction any value of $(c_1,c_2,c_3,c_4)$ will only activate the normalizable modes and hence retain the correct asymptotic form.  Moreover, for any generic choice of $c_i$'s determining a decreasing 10d dilaton, we always flow to a smeared D2 brane singularity. However though, some special constraints must be imposed on these parameters in order to have a correct intermediate bubble regime, thus realizing the aforementioned gluing conditions.

Indeed, by taking a look at the form of the intermediate bubble solution expressed in the gauge \eqref{D2_gauge} as it appears in the above table, it becomes clear that the field combination $\left(3U-\f{\phi}{4}\right)$ is constant. Hence, if we want this to match the linearized expansion coming from the near-boundary analysis at a somewhat larger $r$, the following linear constraints have to be satisfied
\begin{equation}\label{Lin_BoN_conditions}
\begin{array}{lccclc}
3c_1-4c_2 \ \overset{!}{=} \ 0 & , & & & c_3+12c_4 \ \overset{!}{=} \ 0 & .
\end{array}
\end{equation}
After imposing the above constraints, the left-over freedom amounts to two normalization parameters that will eventually fix the position of the source $r_{\mathrm{D}2}$ and the effective radius of the bubble $R$.\footnote{Given that the initial perturbation satisfies $\delta\phi>0$, the numerical integration for any value of these remaining free parameters will result in a qualitatively identical result.} 

\begin{figure}[!htb]
	\centering
	\includegraphics[width=0.98\textwidth]{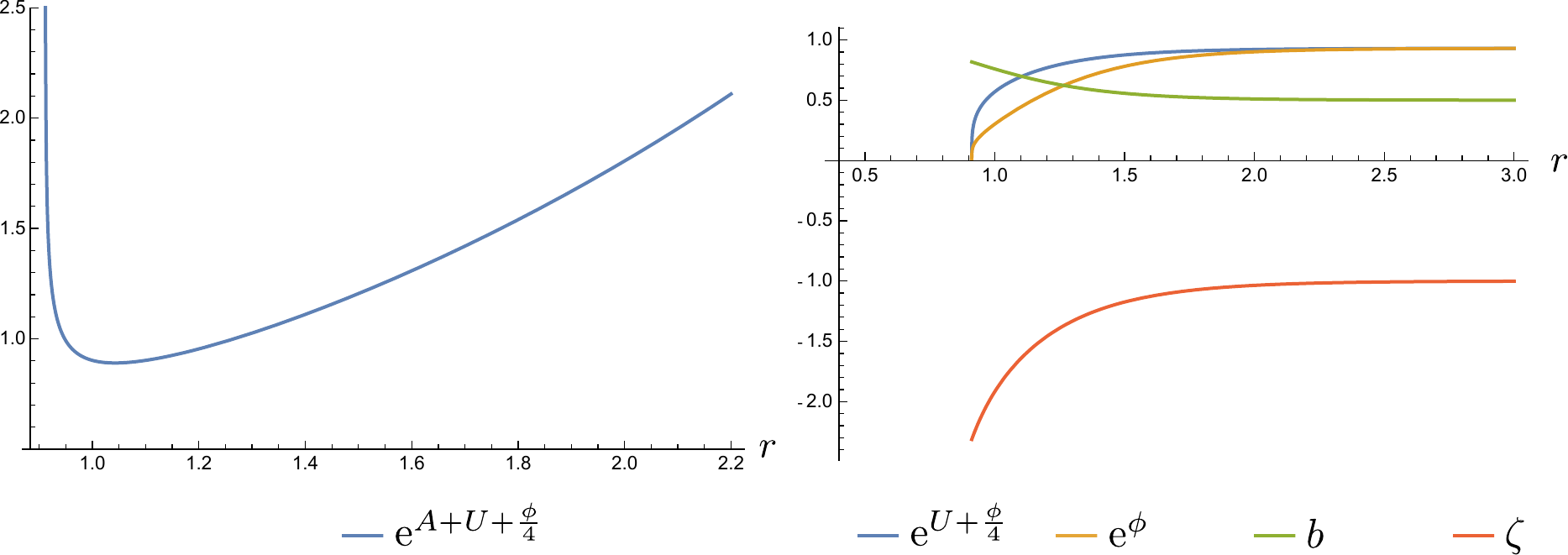}
	\caption{\it The explicit radial flow describing the instability of the non-SUSY G$_2$ vacuum. The $\e^{A+U+\frac{\phi}{4}}$ and $\e^{U+\frac{\phi}{4}}$ warp factors respectively control the size of $\dS_3$ and $S^6$ in the string frame metric (cf.~\eqref{10ansatz}, recalling that $V=U+\frac{\phi}{4}$). These ones, together with the remaining 10d fields are plotted as functions of $r$.}
	\label{fig:DbubblePlot}
\end{figure}

The explicit choice made here is 
\begin{equation}
	\begin{aligned}
		c_1 &= -2(3+\sqrt{33})\,, \qquad\qquad & c_2 &=  -\f32 (3+\sqrt{33})\,, \\
		c_3 &= \f32(3+\sqrt{89})\,, & c_4 &= -\f18(3+\sqrt{89})\,.
	\end{aligned}
\end{equation}
The result of this integration is given in Figure \ref{fig:DbubblePlot}. 

\begin{figure}[!htb]
	\centering
	\includegraphics[width=0.75\textwidth]{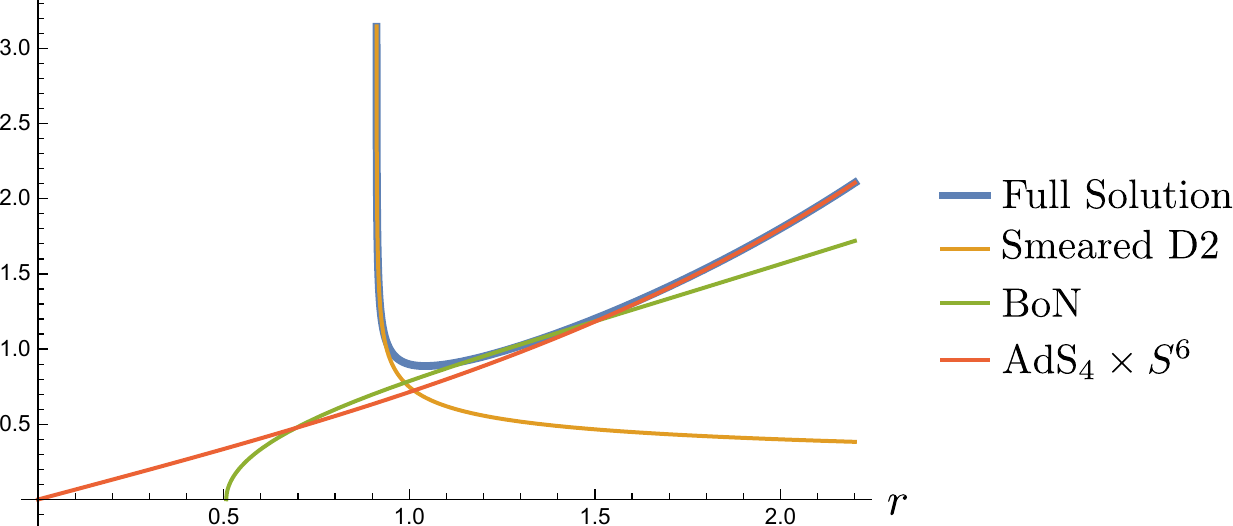}
	\caption{\it The radial profile of the warp factor $\e^{A+U+\f{\phi}{4}}$ in front of $\dS_3$ in the 10d string frame metric (blue curve). The actual flow interpolates between a source regime (yellow curve) and the asymptotic AdS regime (red curve), going through an intermediate bubble regime (green curve). Similar plots are represented below  for the other functions controlling the solution.}
	\label{fig:Regimes_U}
\end{figure}

\begin{figure}[!htb]
	\centering
	\includegraphics[width=0.75\textwidth]{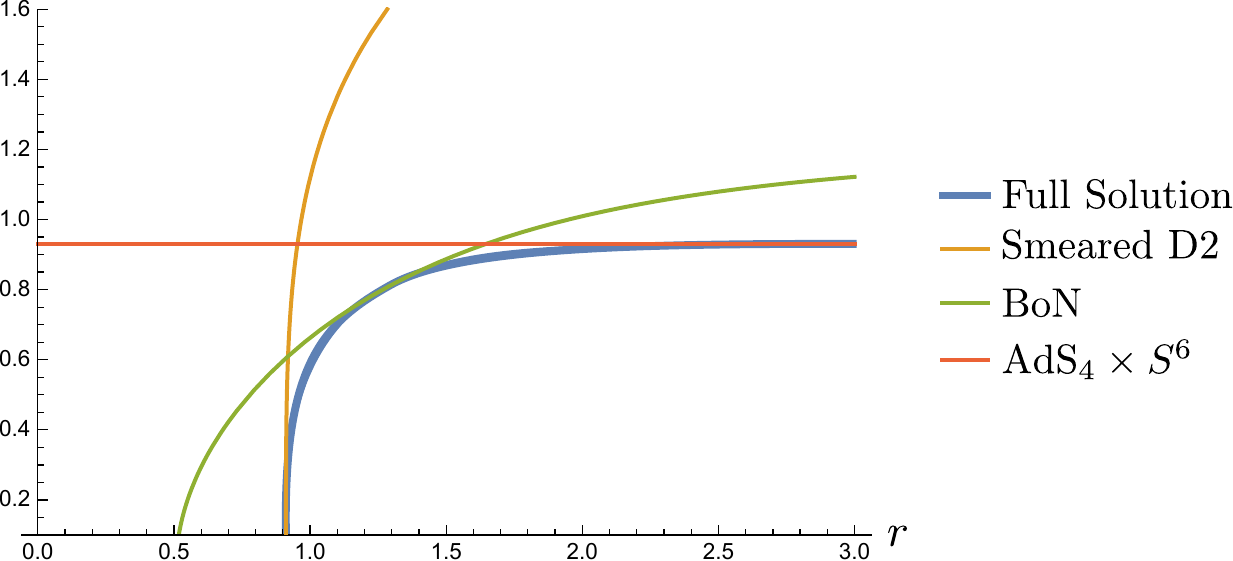}
	\caption{\it  The radial profile of the warp factor $\e^{U+\f{\phi}{4}}$ controlling the size of $S^6$ in the 10d string frame metric.
	}
	\label{fig:Regimes_W}
\end{figure}

\begin{figure}[!htb]
	\centering
	\includegraphics[width=0.75\textwidth]{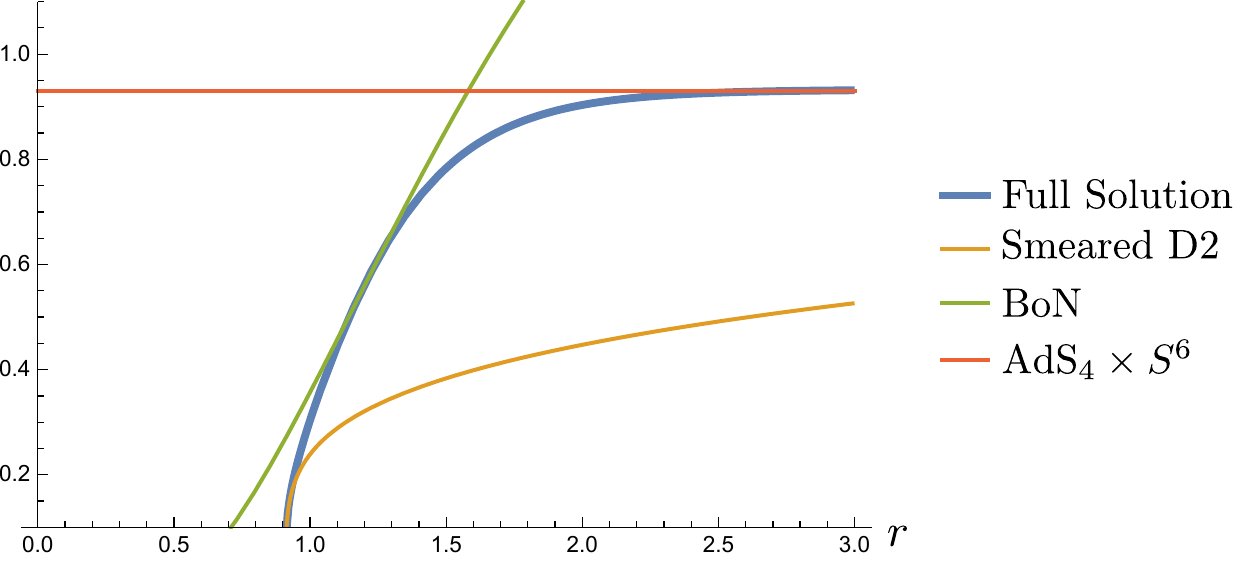}
	\caption{\it  The radial profile of the 10d dilaton $\e^{\phi}$.
	}
	\label{fig:Regimes_phi}
\end{figure}

Finally, Figures \ref{fig:Regimes_U}, \ref{fig:Regimes_W} and \ref{fig:Regimes_phi} show in more detail how the various fields in our solution interpolate among the three different regimes that we have described above.
%

\section{Discussion}
\label{sec:discussion}
In this paper we constructed numerical bounce solutions in massive type IIA supergravity and argued that they represent a non-perturbative decay channel for its non-supersymmetric ${\rm G}_2$ invariant $\AdS_4\times S^6$ vacuum solution. These instabilities also apply to related $\AdS_4\times \cM_6$ solutions, where $\cM_6$ has a nearly-K\"ahler structure. Analogous to \cite{Horowitz:2007pr}, these bounce solutions describe a type of bubble instability of the vacuum. These are however not the vanilla type BoN's as introduced by \cite{Witten:1981gj}, as 
 the presence of a Freund-Rubin type $F_6$ flux on $S^6$ prevents such bubbles from persisting all the way through the flow. Instead they are BoN's dressed up with D2 branes extending along $\dS_3$ and homogeneously distributed over~$S^6$. 

\textbf{Instability of the non-supersymmetric vacuum:} For the above reasons, the physical interpretation of the actual decay channel for the non-supersymmetric G$_2$ vacuum is rather the mechanism of spontaneous nucleation of spherical D2 bubbles, in line with an Ooguri-Vafa instability of the type discussed in \cite{Ooguri:2016pdq}. It is worth mentioning that, since the supergravity approximation is expected to be unreliable in the neighborhood of the source, one cannot simply evaluate the on-shell Euclidean action in order to estimate the decay rate. This is very different than the situation of a typical semiclassical instanton geometry.

Nevertheless, a possible way of estimating this is to follow~\cite{Horowitz:2007pr} and combine the Euclidean action of the pure bubble background introduced in Section~\ref{Summary} with the action of $N$ probe D2 branes placed at the location of the bubble, i.e.\ at $\rho = R$, where $N$ has to be chosen so that it matches the Page $F_6$ flux of the AdS$_4\times \mathcal{M}_6$ vacuum. Both these contributions scale like $R^3$. The value of this parameter will eventually be determined by the parameters of the full numerical solution.

One can also check that the patch identified by the de Sitter slicing of AdS$_4$, which contains our bubble solution, reaches the boundary in finite global time.

 \textbf{Dynamical obstruction for the supersymmetric vacuum:} Numerical flow solutions similar to the one presented above are quite generic. Indeed we find that within the truncation  studied above, qualitatively similar solutions can be constructed that asymptote either to the non-supersymmetric, perturbatively unstable SO$(7)$ invariant vacuum, or to the supersymmetric vacuum of \cite{Behrndt:2004km}. The existence of such a flow asymptotic to the supersymmetric AdS vacuum is puzzling, as these vacua are expected to be stable by virtue of positive energy arguments (see e.g.~\cite{Gibbons:1983aq}).
 Although we do not have a definite answer to this puzzle, we notice
that the solution analog to the one above, but asymptotic to the supersymmetric vacuum, turns out not to have the appropriate intermediate regime that makes it naturally interpreted as a bubble. In order to see this, one can start by looking at the asymptotic behavior of the solution, namely at the linearized expansion around the supersymmetric vacuum. The mass eigenstates here are not organized pairwise, and as a consequence, the eigenspaces of linearized normalizable modes are all one-dimensional. Hence, trying to impose gluing conditions such that $\left(3U-\f{\phi}{4}\right)$ is constant will completely fix the coefficients as $c_1=c_2=c_3=c_4=0$ (instead of giving \eqref{Lin_BoN_conditions}), which just corresponds to empty AdS$_4$. On the other hand, without such constraints one obtains solutions displaying an intermediate regime that is still well approximated by \eqref{BoN_Schw_gen}, but with $\alpha\neq\f{1}{2}$, so it does not match the bubble geometry \eqref{BoN_Schw_gauge}.

In fact we find that the solution asymptotic to the supersymmetric vacuum and the one asymptotic to the non-supersymmetric vacuum discussed in Section~\ref{sec:numerics} have different features in the interior.
This can be seen by considering a probe spherical D2 brane in a background of the form~\eqref{10ansatz}, where  $\dS_3$ is now Wick-rotated to $S^3$. The D2 brane wraps $S^3$ and is placed at some radial distance $r$.
The Euclidean DBI+WZ action is given by
\begin{equation}
\begin{aligned}
S_{\mathrm{D2}}(r) \, &= \, \tau_{\mathrm{D}2}\int_{S^3}\dd^3\xi\,\e^{-\tilde{\Phi}}\sqrt{\det\,\tilde{g}_3} - \tau_{\mathrm{D}2}\int_{S^3} \tilde{C}_3\,,
\end{aligned}
\end{equation}
where a tilde denotes the pull-back on the brane of the bulk fields (and we have reabsorbed a factor of $i$ in the three-form).
The force acting on the D2 is then
\begin{equation}
\mathcal{F} = - \frac{\dd}{\dd r} S_{\rm D2}\,.
\end{equation}
Plugging the ansatz~\eqref{10ansatz} in, we obtain
\begin{equation}
\mathcal{F} \,=\, -2\pi^2\tau_{\rm D2}L^3  \left[ \left(\e^{3A+3V-\phi}\right)' + g^6  \e^{3A-2V} f_6 \right]\,,
\end{equation}
where $f_6$ was given in \eqref{bianchifluxes}.
We evaluated this force numerically for the flow asymptotic to the supersymmetric vacuum as well as for the one asymptotic to the non-supersymmetric vacuum discussed in Section~\ref{sec:numerics}. In both cases we find that when $r$ is large the force is attractive towards the interior, in agreement with the absence of brane-jet instabilities found in \cite{Guarino:2020jwv}. We also find that in both cases the force remains finite when approaching the location of the stack of backreacting D2's in the source regime. In particular, we do obtain the cancellation of divergences between the DBI and WZ terms that is expected for mutually BPS objects. However, for the solution connected to the supersymmetric vacuum the force is always attractive towards the interior and non-vanishing, even when the probe brane gets close to the backreacting D2's; this may mean that we have to modify the IR picture of the solution. For the solution asymptotic to the non-supersymmetric vacuum instead, the force is repulsive close to the backreacting D2's, so that the probe has an equilibrium point at finite distance.
 
\textbf{Localized bounces:} One can ask whether in addition to the smeared instantons constructed here, there also  exist localized instantons, that is instantons localized both in EAdS and in $S^6$. Certainly, given the lower amount of symmetry of the localized configurations, these are expected to be hard to construct. In \cite{Horowitz:2007pr} a similar issue is discussed in the context of IIB supergravity on AdS$_5\times S^5/\mathbb{Z}_k$ and localized instantons are argued to exist, although no complete construction is given.  It is also argued there that such solutions should have a smaller action than the smeared instantons. One might expect a similar argument to apply to our case.

\textbf{Generalizations:} In \cite{Guarino:2020flh,Bobev:2020qev} six additional perturbatively and brane-jet stable $\AdS_4$ vacua are described. These additional vacua fall outside the ansatz we describe in \eqref{10ansatz} or equivalently outside of the four-dimensional NK truncation. For this reason our numerical analysis cannot immediately be extended to these cases and we would need to introduce a more general ansatz with additional scalar fields. However, it is likely that the instabilities described in this work are very generic and will also show up in these cases. Indeed, for the (perturbatively unstable) non-supersymmetric $\AdS_4$ solution (with SO($7$) invariance) we found the same type of instability where we saw that the metric functions as well as the dilaton behave essentially identically to the G$_2$ invariant case. The only difference consists in the additional scalars that are present in the latter case. It would be interesting to explicitly construct these in the adequate consistent truncation and show that indeed all non-supersymmetric solutions described in \cite{Guarino:2020flh} exhibit the same non-perturbative instabilities.

To conclude, it may be worth stressing that the novel dilaton bubbles constructed in this paper might play a much more general role in establishing the non-perturbative instability of string theory AdS vacua. This is due to the fact that these objects evade universal obstructions concerning the possibility of having a shrinking internal space in the presence of flux.
In particular, it would be extremely interesting to test the possible relevance of these bubbles in various other stringy settings, including type IIB/F-theory setups like the ones underlying the novel non-supersymmetric S-fold constructions of \cite{Guarino:2021hrc}.

\bigskip
\bigskip
\leftline{\bf Acknowledgments}
\smallskip
\noindent We are grateful to Luca Martucci for useful discussions and comments on the manuscript. We would also like to thank Irene Valenzuela for useful correspondence. The work of PB and GD is supported by the STARS-StG grant THEsPIAN. PB is supported by Fellowships of the Belgian American Educational Foundation and Fulbright. The work of NP is supported by the Principado de Asturias through the grant FC-GRUPIN-IDI/2018/000174.


\appendix
	
\section{Supergravity conventions}
\label{app:sugra}

The bosonic field content of massive type IIA supergravity consists of a metric $G_{MN}$, a dilaton $\Phi$, the NS-NS three-form $H_3$ and R-R $n$-form field strengths $F_n$, with $n=0,2,4$. The Romans mass, $F_0$, does not have any propagating degrees of freedom. Since we are only interested in bosonic solutions we will not need to discuss the fermionic sector. 

We use the democratic formalism in which the number of R-R fields is doubled such that $n$ runs over $0,2,4,6,8,10$ \cite{Bergshoeff:2001pv}.\footnote{We use the conventions of~\cite{Koerber:2010bx}, albeit with $B_2^{\rm there}= - B_2^{\rm here}$.} This redundancy is removed by introducing duality conditions for all R-R fields
\begin{equation}
	F_n = (-1)^{\f{(n-1)(n-2)}{2}}\star_{10} F_{10-n} \,.
\end{equation}
These duality conditions should be imposed by hand after deriving the equations of motion from the action. We define the NS-NS and R-R gauge potentials as follows
\begin{equation}
H_3 = \dd B_2\,,
\end{equation}
\begin{equation}\label{fieldstrengths}
	F_n = \dd C_{n-1} - H_3\wedge C_{n-3} + F_0\, \e^{B_2}\big|_n\,,
\end{equation}
where $\e^{B_2}\big|_n$ denotes the degree $n$ term in the expansion of the exponential.

The bosonic part of the action in string frame is 
\begin{equation}\label{actionII}
	S_{{\rm bos}} = \f{1}{2\kappa_{10}^2}\int \star_{10}\Big[ \e^{-2\Phi}\Big( R + 4 |\dd \Phi|^2 - \f{1}{2} |H_3|^2 \Big) - \f{1}{4} \sum_{n}|F_n|^2 \Big]\,,
\end{equation}
where the ten-dimensional Newton's constant $\kappa_{10}$ is related to the string length through $4\pi\kappa_{10}^2=(2\pi l_s)^8$ and the Hodge star is defined such that for any $n$-form $A$ one has
\begin{equation}
	\star_{10}|A|^2 \,\equiv\, \star_{10}\f{1}{n!} A_{M_1\dots M_n}A^{M_1\dots M_n} = \star_{10} A \wedge A \,.
\end{equation}
The Bianchi identities and equations of motion derived from the action \eqref{actionII} are
\begin{equation}
	\dd H_3 =0\,, \qquad \text{and}\qquad 
	\dd (\e^{-2\Phi}\star_{10} H_3) + \f{1}{2}\sum_n \star_{10} F_n\wedge F_{n-2} =0\,,
\end{equation}
for the NS-NS field $H_3$ and
\begin{equation}\label{Bianchi10d}
	\dd F_n -H_3 \wedge F_{n-2} = 0\,,
\end{equation}
for the R-R form fields. The dilaton and the Einstein equations of motion can be written as
\begin{equation}
	\begin{aligned}
		0 &= \nabla^2 \Phi - |\dd \Phi|^2 + \f14 R - \f18 |H_3|^2  \, ,\\
		0 &= R_{MN} + 2 \nabla_M\nabla_N \Phi - \f12 |H_3|^2_{MN} - \f14 \e^{2\Phi} \sum_n |F_n|^2_{MN}  \,,
	\end{aligned}
\end{equation} 
where we have defined
\begin{equation}
	|A_n|^2_{MN} \equiv \f{1}{(n-1)!}{(A_n)_{M}}^{M_2\cdots M_n}(A_n)_{NM_2\cdots M_n} \,.
\end{equation}

\section{Nearly K\"ahler structure and the six-sphere}
\label{app:S6}
 
A six-dimensional nearly-K\"ahler manifold has $\SU(3)$ structure group and can thus be equipped with a real two-form $J$ and a complex decomposable three-form $\Omega$. These forms are of type $(1,1)$ and $(3,0)$ with respect to the natural almost complex structure defined by $\Omega$. They are subject to the defining conditions:
\begin{align}
	\Omega \wedge \overline{\Omega} &= - \f{4i}{3}J\wedge J\wedge J\,,& J\wedge \Omega &=0 \label{alg} \\
	\dd\, \Im\,\Omega &= - 2J\wedge J\,,& \dd J &= 3\,\Re\,\Omega\label{dif}\,.
\end{align}
The volume form is $\star_6 1 = {\rm vol}_6 = \f{1}{6} J\wedge J \wedge J$, hence $\star_6 J = \f{1}{2} J\wedge J $ and $\star_6 \Omega =  i\,\Omega$.

The cone $C[\cM_6] = \mathbf{R}_+ \times \cM_6$ has G$_2$ holonomy and thus it allows for the existence of an associative three-form $\psi$ and a co-associative four-form $\wti\psi = \star_7 \psi$ satisfying $\dd \psi = \dd \wti\psi =0$ (the Hodge star $\star_7$ is understood to use the metric $\dd s^2_7 = \dd r^2 + r^2 \dd s^2_{6}$ on the cone $C[\cM_6]$). 
The 6d SU(3) structure and the 7d G$_2$ structure are related as
\begin{equation}\label{psJOm}
	\psi = r^2\,\dd r\wedge J + r^3 \,\Re \Omega\,,\qquad \widetilde{\psi} = \f12 r^4 J\wedge J - r^3 \dd r\wedge \Im \Omega\,,
\end{equation}
and the differential constraints \eqref{dif} are equivalent to closure of the (co-)associative forms. 

The complete list of homogeneous six-dimensional nearly-K\"ahler manifolds is given by $S^6$, $\mathbb{CP}^3$, $S^3\times S^3$, and the flag manifold $\mathbb{F}^3$~\cite{Butruille}, while non-homogeneous examples have been found in \cite{FoscoloHaskins}. 

Finally we discuss how the nearly-K\"ahler structure is constructed on the six-sphere $S^6$ with G$_2$ invariance.
To make this explicit we define the embedding coordinates $\mu^I$ parameterizing the unit radius six-sphere $S^6$ as the locus
\begin{equation}
	\sum_{I=1}^7 \left(\mu^I\right)^2 =1
\end{equation}
in $\mathbf{R}^7$. When we take $S^6={\rm G}_2/\SU(3)$, the $\SU(3)$ structure is invariant under the transitive action of ${\rm G}_2$ and hence we can specify it in terms of the embedding of $S^6$ in it's cone. To this end let us introduce the unconstrained coordinates $x^I = r \mu^I$ on the cone $C[S^6] = \mathbf{R}^7$, with the associated orthonormal siebenbein $e^A_I$. We can then define the (co-)associative three- and four-form as 
\begin{equation}
\begin{aligned}
	\psi &= e^{123}+e^{145}-e^{167}+e^{246}+e^{257}+e^{347}-e^{356}\,,\\[1mm]
		\tilde{\psi} &= e^{4567}+e^{2367}-e^{2345}+e^{1357}+e^{1346}+e^{1256}-e^{1247}\,,
\end{aligned}
\end{equation}
where $e^A = e_I^A \dd x^I$ and $e^{ABC} = e^A\wedge e^B\wedge e^C$. Comparing this with \eqref{psJOm}, we find an expression of the nearly-K\"ahler forms on $S^6$ in terms of the (co-)associative forms on $\mathbf{R}^7$ as
\begin{equation}
\begin{aligned}
	J &= \iota_{\partial_r} \psi\big|_{r=1} = \f12 \psi_{IJK} \mu^I \dd \mu^J\wedge \dd \mu^K\,,\\ 
	\Omega &= \psi\big|_{r=1} - i \,\iota_{\partial_r} \tilde{\psi}\big|_{r=1} = \f16 \left(\psi_{JKL}-i \tilde{\psi}_{IJKL}\,\mu^I\right) \dd \mu^J\wedge \dd \mu^K\wedge \dd \mu^L\,.
\end{aligned}
\end{equation}
One can check that these forms indeed satisfy the relations in \eqref{alg} and \eqref{dif}. The metric specified by this nearly-K\"ahler structure is exactly the round metric on $S^6$. 

\section{Rescaling the 4d fields}\label{AppRescaling}

 We now show that making suitable rescalings, all dependence on the parameters $m$ and $g$ in the reduced action \eqref{1dAction} is moved to just an overall prefactor. This allows us to easily relate the choice $m=1$, $g=1/2$ made in the main text to any other choice.
 
We start by trading $g$ for the dimensionless parameter 
\begin{equation}\label{defmu}
\mu=\left(\f{2g}{m}\right)^{1/3}\,.
\end{equation}
Next we introduce a new (dimensionless) radial coordinate $\ti r$ through
\begin{equation}\label{rescaling_radial_coord}
r=m^{-1}\mu^{-3}\,\tilde r\,,
\end{equation}
we rescale the scalar fields as
\begin{equation}
		b(r) = \mu\, \ti{b}(\ti r)\,,\quad  \zeta(r) = \mu^{-1}\, \ti{\zeta}(\ti r)\,,\quad \e^{\phi(r)} = \mu^{5/2}\, \e^{\ti\phi(\ti r)}\,,\quad \e^{U(r)} = \mu^{-1/8}\,\e^{\ti{U}(\ti r)}\,,
\end{equation}
and the 4d metric as
\begin{equation}
\dd \tilde{s}^2_4 \,=\, m^2\mu^7 \dd s_4^2 \,=\,  \e^{8\ti U(r)}\big( \dd \ti r^2 + \tilde{L}^2\,\e^{2\ti A(\ti r)} \ds^2_{3} \big)\,,
\end{equation}
with 
\begin{equation}\label{lastrescaling}
A(r) \,=\, \ti{A}(\ti r)\,,\qquad L \,=\, m^{-1}\mu^{-3} \ti{L}\,.
\end{equation}
Note that the new metric takes the same form as \eqref{4dmetric} in the rescaled variables.

In these variables, the reduced action \eqref{1dAction} reads
\begin{equation}
\begin{aligned}\label{1dAction_rescaled}
	S_{\rm 1}&= \f{\ti{L}^3\vol_3}{2m^2\mu^7\kappa_4^2} \int \dd \ti r\left[ \e^{3\ti A+8\ti U}\left( 6 (\ti A'+4\ti U')^2 - 24 (\ti U')^2 - \f12(\ti\phi')^2 - \f32\e^{-\ti\phi-4\ti U}(\ti b')^2\right. \right.\\[1mm]
		&\hspace{5.8cm}\left.\left. -\f12\, \e^{\f{\ti\phi}{2}-6\ti U}(\ti\zeta')^2 \right)
 	 \ - \,\e^{3\ti A+16 \ti U}\ti{\mathcal{V}} + \f{6}{\ti{L}^2}\,\e^{\ti A+8\ti U} \right]   \,,
\end{aligned}
\end{equation}
where now the prime denotes a derivative with respect to $\ti r$, and the rescaled potential is
\begin{equation}
\begin{aligned}
\ti{\mathcal{V}} \,=\, m^{-2}\mu^{-7} \mathcal{V} \,&=\, \f12 \left[ \e^{-\f{\ti{\phi}}{2}-18\ti{U}}\big(\,\tfrac{5}{2} + \ti  b^3 + 3 \ti b\ti\zeta\,\big)^2 + 3\,\e^{\f{\ti\phi}{2}-14\ti U}(\ti b^2+\ti \zeta)^2\right. \\ 
&\ \qquad\left.-15\,\e^{-8\ti U}+9\ti b^2 \, \e^{-\ti\phi-12\ti U} + 3\ti b^2\, \e^{\f32\ti\phi-10\ti U} + \e^{\f52\ti\phi-6\ti U}\right]\,.
\end{aligned}
\end{equation}
Now the parameters of the theory appear in the action only through an overall prefactor. 

These manipulations allow us to infer that the choice $m=1, g=1/2$ made in the main text does not entail a loss of generality. Indeed, for $m=1, g=1/2$ (which implies $\mu=1$), the original variables coincide with the tilded variables introduced here, and the same holds for the respective equations of motion. This means that the analysis in the main text is the same as an analysis done using the tilded variables.  Hence the solution discussed there can be straightforwardly converted into a  solution for any other value of $m,g$ by implementing the rescalings \eqref{rescaling_radial_coord}--\eqref{lastrescaling}. In particular, the numerical solutions we construct in the main text using $m=1, g=1/2$ would look qualitatively the same (modulo the rescalings) for any other choice of the parameters.

\bibliography{BoN}
\bibliographystyle{JHEP}
	
\end{document}